# Liquefaction-induced Plasticity from Entropy-boosted Amorphous Ceramics

Haidong Bian[1,2,3,4], Quanfeng He[4], Junhua Luan[4], Yu Bu[2,4], Yong Yang[1,3,4,*], Zhengtao Xu[5], Jian Lu [1,3,4,6,*], Yang Yang Li [1,2,3,4,6,*]

[1] Hong Kong Branch of National Precious Metals Material Engineering Research Centre, City University of Hong Kong, Tat Chee Avenue 83, Kowloon, Hong Kong, China

[2] Center of Super-Diamond and Advanced Films (COSDAF), City University of Hong Kong, Tat Chee Avenue 83, Kowloon, Hong Kong, China

[3] Department of Materials Science and Engineering, City University of Hong Kong, Tat Chee Avenue 83, Kowloon, Hong Kong, China

[4] Department of Mechanical Engineering, City University of Hong Kong, Tat Chee Avenue 83, Kowloon, Hong Kong, China

[5] Department of Chemistry, City University of Hong Kong, Tat Chee Avenue 83, Kowloon, Hong Kong, China

[6] Centre for Advanced Structural Materials, City University of Hong Kong Shenzhen Research Institute, Greater Bay Joint Division, Shenyang National Laboratory for Materials Science, 8 Yuexing 1st Road, Shenzhen Hi-Tech Industrial Park, Nanshan District, Shenzhen, China

E-mail: yonyang@cityu.edu.hk; jianlu@cityu.edu.hk; yangli@cityu.edu.hk

Highlights

- Amorphous oxides are fabricated by anodization of mid-entropy alloys
- The amorphous ceramics can withstand severe plastic deformation over 95%.
- The high plasticity of the ceramics is due to the stress-triggered liquefaction.

**Abstract**

Ceramics are easy to break, and very few generic mechanisms are available for improving their mechanical properties, e.g., the 1975-discovered anti-fracture mechanism is strictly limited to zirconia and hafnia. Here we report a general mechanism for achieving high plasticity through liquefaction of ceramics. We further disclose the general material design strategies to achieve this difficult task through entropy-boosted amorphous ceramics (EBACs), enabling fracture-resistant properties that can withstand severe plastic deformation (e.g., over 95%, deformed to a thickness of a few nanometers) while maintaining high hardness and reduced modulus. The findings reported here open a new route to ductile ceramics and many applications.

## 1. Introduction

Ceramics are hard but brittle materials, and it remains a grand challenge to enhance their plasticity for more versatile applications.[1] Current solutions mainly depend on compositing with foreign materials, such as polymers,[2-5] carbon,[6-9] and metals.[8,10-14] Other research efforts include metallographic approaches (e.g., through creating grained or twinned microstructures in ceramics[15-17]) and smart designs of elaborate ceramic microarchitectures, generally entailing complex and expensive nano-machining techniques and restricted to small sample sizes.[5,18,19] However, for pure ceramics, there are very few fracture-resistance mechanisms available. For example, the phase-transformation mechanism discovered in 1975[20] is only applicable to zirconia and the high-temperature phase of hafnia.[21] Specifically, the peculiar transformation-toughening mechanism[22-24] involves the phase transformation from the metastable tetragonal phase to the monoclinic phase at the crack tip under a concentrated external stress, leading to significant volume expansion that compresses the crack and prohibits its growth. Another very recent study reported the very high plasticity achieved on glassy $Al_2O_3$ films, but requires flawless film quality.[25] Limited progresses aside, the making of plastic and tough ceramics remains an outstanding important target for modern technology.

We here disclose a generic mechanism for enhancing both flowability and plasticity in ceramics through stress-induced liquefaction, and further propose the usage of entropy-boosted amorphous ceramics (EBACs) for accomplishing this challenging task. Although "amorphous" and "glassy" are sometimes synonymously used, strictly speaking, "amorphous" is not interchangeable with "glassy", as the latter specifically refers to the non-crystalline materials obtained by quenching the parent liquid, particularly in the fields of metallurgy and solid state matters. Therefore, to avoid confusion, we will not use the common term of "glass transition" for describing the amorphous to liquid transition--instead, we will use "liquefaction".

## 2. Results and discussion

Anodic (TiAlV)$O_x$ films were studied as a demonstrative material system of EBACs. Amorphous (TiAlV)$O_x$ (Figures 1d and S1) was fabricated by anodizing the parent entropy-stabilized alloy of TiAlV. Anodization is a convenient solution-based technique that is particularly suitable for producing amorphous oxides.[26,27] The as-anodized (TiAlV)$O_x$ films exhibited a smooth surface to naked eyes with a rich variety of vivid hues (Figure 1a), while possessing uniform mesopores at the microscopic scale (Figures 1b, S2-S3). Taking the (TiAlV)$O_x$ anodized at 100 V for 2 hrs as a representative, the film displayed an average pore size of 23 nm, ligament thickness of 27 nm, porosity of 55%, pore depth over 55 nm, and film thickness of 260 nm (Figures S2-S5, Table S1). According to the elemental mapping and depth profile analysis, the anodic (TiAlV)$O_x$ films featured highly uniform elemental distribution (Figures 1e-f and S6-S7). The above observations show that the anodic (TiAlV)$O_x$ is an amorphous solid solution of multi-cationic oxides, representing a novel type of EBACs.

Based on the Olive and Pharr method,[28] the reduced modulus ($E_r$) and hardness ($H$) of the as-anodized (TiAlV)$O_x$ films were estimated using the indentation tests (measured at the displacement depth of 50 nm): $E_r(H)$ = 192.5(13.2) GPa (Figure S4a and Table S1), outperforming the TiAlV substrate which displayed a $Er(H)$ of 146(11.2) GPa (Figure S8). Note that the pore area was not excluded in the material contact area in the above calculation, thus the $Er$ and $H$ reported here for the porous (TiAlV)$O_x$ films were actually under-estimated.[29] Interestingly, Figure S4b shows that the reduced modulus of the porous (TiAlV)O$_x$ film remains almost to be a constant (only slightly decreases at the beginning of indentation). This behavior is in sharp contrast to typical plasticity in porous materials[30], which generally results in densification and therefore increasing modulus with the indentation depth. In other words, the mechanism of densification-dominated plasticity can be ruled out for our porous (TiAlV)O$_x$ film.

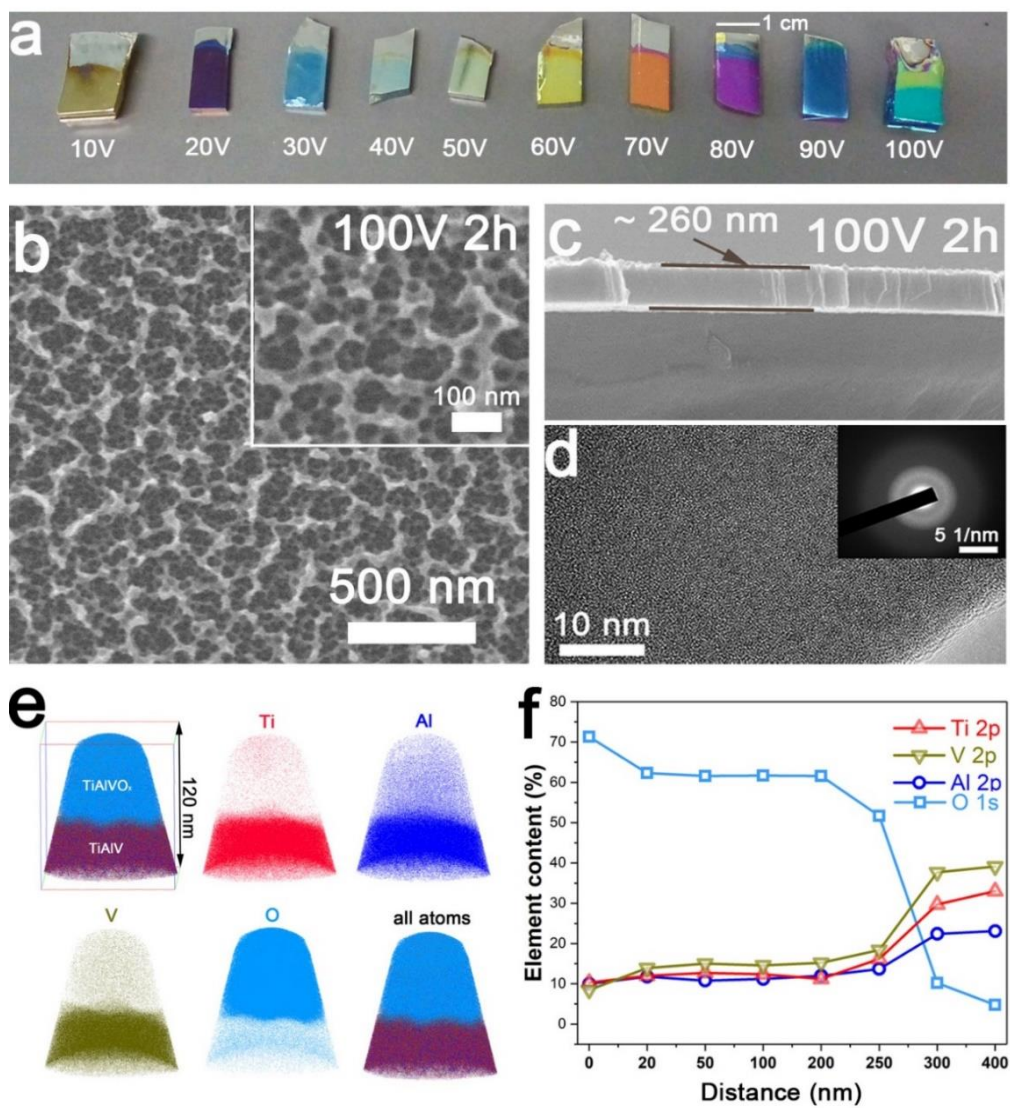

**Figure 1.** Optical photographs (a) of the as-anodized (TiAlV)O$_x$ films that were synthesized in 0.1 M oxalic acid at different potentials (10 ~ 100 V) for 2 hrs, top-view (b) and cross-sectional (c) SEM images, and the corresponding HRTEM image (d), SAED pattern (inset of d) of the sample anodized at 100 V for 2 h. 3D ATP reconstruction map with elemental mapping (e) and XPS elemental depth-profiles (f) of the anodic (TiAlV)O$_x$ film grown on the TiAlV substrate (produced by anodization at 100 V for 2 hrs).

Ceramics are known to fracture easily once stressed beyond its yield strength, unable to endure severe plastic deformation at the crack region. It should be pointed out that, besides the direct measurements using tensile tests, plasticity is often indirectly investigated using the indentation method, particularly for non-free standing materials (e.g., thin films or coatings) that cannot be directly tested under tension.[31,32] For the (TiAlV)O$_x$ film, plasticity could be estimated by measuring the ratios of the residue displacement to the maximum displacement on the nanoindentation load/unload curves (Figure 2, a and e), showing values close to 61%.[33,34] SEM images of the indents pressed with different loading forces (500 μN ~ 5 N) (Figure 2) further verified the films' exceptional tolerance to plastic deformation. Although the mesoporous film was severely stretched and deformed upon indentation, no superficial cracks or detachments were detected at the indented area, even under the equipment's maximum load of 5 N (Figure 2). Moreover, shear bands, which are normally only found on ductile metallic materials as a symbol of severe plastic deformation, were observed after indentation (Figure 2, h-l), again evidencing the film's high plasticity. By comparison, conventional ceramics are normally found to be brittle and fragile upon indentation. For example, when compressed under a force a magnitude smaller than 5 N, mica single crystals and graphite were observed to exhibit obvious spallation, pile-ups, cracking, and damages, indicating their low tolerance to external stress.[35] Moreover, it is worth mentioning that porous films are often observed to exhibit stripping or cracks around the imprint upon indentation especially under a high load (such as those used here).[36,37]

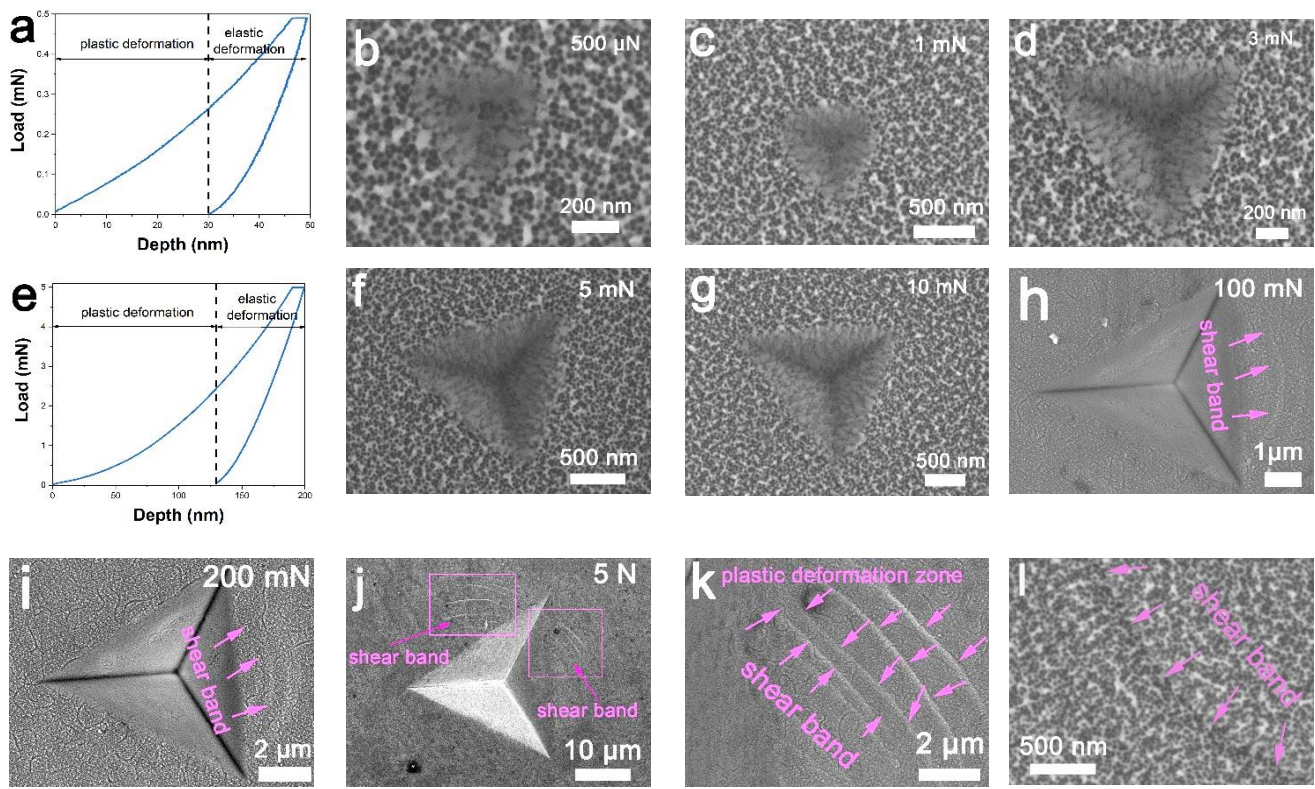

**Figure 2.** The load-unload displacement curve of nanoindentation tests on the (TiAlV)O$_x$ film (anodized at 80 V for 2 h) under different load: a) 500 μN and e) 5 mN. SEM images of nanoindentation impression on the surface of (TiAlV)O$_x$ films at different loadings: b) 500 μN, c) 1 mN, d) 3 mN, f) 5 mN, g) 10 mN, h) 100 mN, i) 200 mN, and j) 5 N. k) and l) are the enlarged SEM images of slip band in j). All the nanoindentation tests were carried out with a loading/unloading time of 5 s and a holding time of 2 s.

The high plasticity of anodic (TiAlV)O$_x$ was further confirmed by the TEM observation of the cross-sectional slices of the indented film (Figures 3 and S9-S11). Note that plasticity could also be evaluated by calculating the percentage change of the film thickness. Therefore, for the film indented with 5 N, the plasticity stretched from 29% at the indent edge (210 and 150 nm thick before and after indentation) to over 95% at the indent tip (less than 10 nm thick after indentation), while the indented film remained crack-free even at the indent tip. Note that this high level of plasticity over 95 % is rarely observed in other ceramic films. Interestingly, the originally nanoporous film appeared to be rather compact and uniform after indentation from the side views. Further examination revealed that the amorphous nature was retained at the indent edge (Figure 3b-c), whereas nanocrystals

(approximately 2 - 5 nm big) emerged (Figure 3d-i) nearby the indent tip under drastic plastic deformation. The inter-plane spacing of the nanocrystals was measured to be 0.209 nm, possibly corresponding to the (402) plane of $Al_2O_3$ (PDF# 50-1496). For confirmation, the sample was fractured on purpose and then scanned with the Raman spectroscope. A new peak at 750 cm$^{-1}$ appeared at the fracture zone, indicating the stress-induced crystallization of metal oxides (Figure 4e). Furthermore, the $(TiAlV)O_x$-coated TiAlV substrate was pulled broken into two pieces, and the fractured regions were examined under the SEM (Figure 4c-d). Clearly the film remained well attached to the substrate without any detachment while displaying highly ductile fracture behaviors.

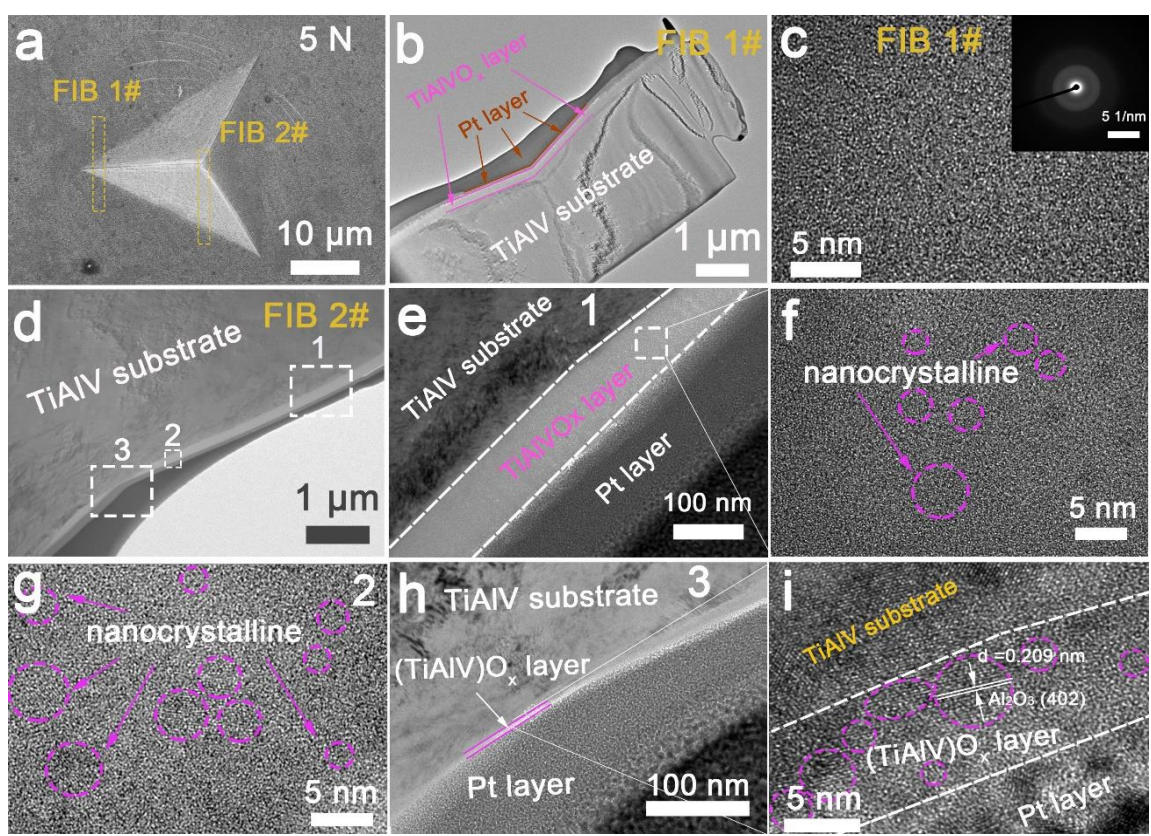

**Figure 3.** SEM image (a) showing the original position of the two FIB-cut slices from the indent area (pressed with 5 N) of the as-anodized $(TiAlV)O_x$ film (anodized at 80 V for 2 hrs) and their corresponding TEM images (b-i). The inset in (c) is the SAED pattern of the anodic film in FIB 1# slice.

The high mechanical performance and crystallization phenomenon are attributed to the stress-induced liquefaction of the EBACs. Some strong evidences include the film morphological change from nanoporous to compact at the indent tip (Figure 3), the smeared doughy surface features inside the indentation (Figure 2), the ductile fracture behaviours (Figure 4 c-d), the emergence of oxide nanocrystals at the indent tip (Figure 3), and, most discolsingly, the "splashed" material on the bank of the indent upon rapid loading and unloading (Figure 4a). Furthermore, the surface features of the $(TiAlV)O_x$ film appeared to be significantly coarsened and coated with a smooth "glaze" (more apparent inside the indented area) upon extended e-beam radiation at 5 KeV in the scanning electron microscope (Figure 4b). This interesting liquefaction phenomenon induced by e-beam heating is unusual for conventional ceramic oxides. All these evidences clearly revealed the amorphous to liquid transformation in the EBACs.

The liquefaction capability of the EBACs is owing to its high level of atomic vacancies, e.g., the anodic $(TiAlV)O_x$ hosts abundant oxygen vacancies as revealed by the split O 1s XPS spectra (Figure S7, d-i). Furthermore, by utilizing the entropy-boosted strategy,[38,39] different species of metal cations and oxygen anions are uniformly distributed in the atomic network, offering a broad scope for atomic inter-binding (as evidenced by the broad XPS features, Figure S7) and high tolerance of bond switching under stress. Therefore, to achieve anti-fracture ceramics, the EBACs appeared to be a promising choice with low steric hindrance and high bond-switching flexibility.

Following this design principle, we further synthesized another new type of EBAC, $(MgZnCa)O_x$, using the sputtering method. The sputtered $(MgZnCa)O_x$ film (Figure S12) also displayed interesting liquefaction phenomena. Remarkably, upon indentation, curious "water stains" were spotted around the fault lines where the stress was concentrated (Figure 4, f-g), implying that the material underwent liquefaction under high stress.

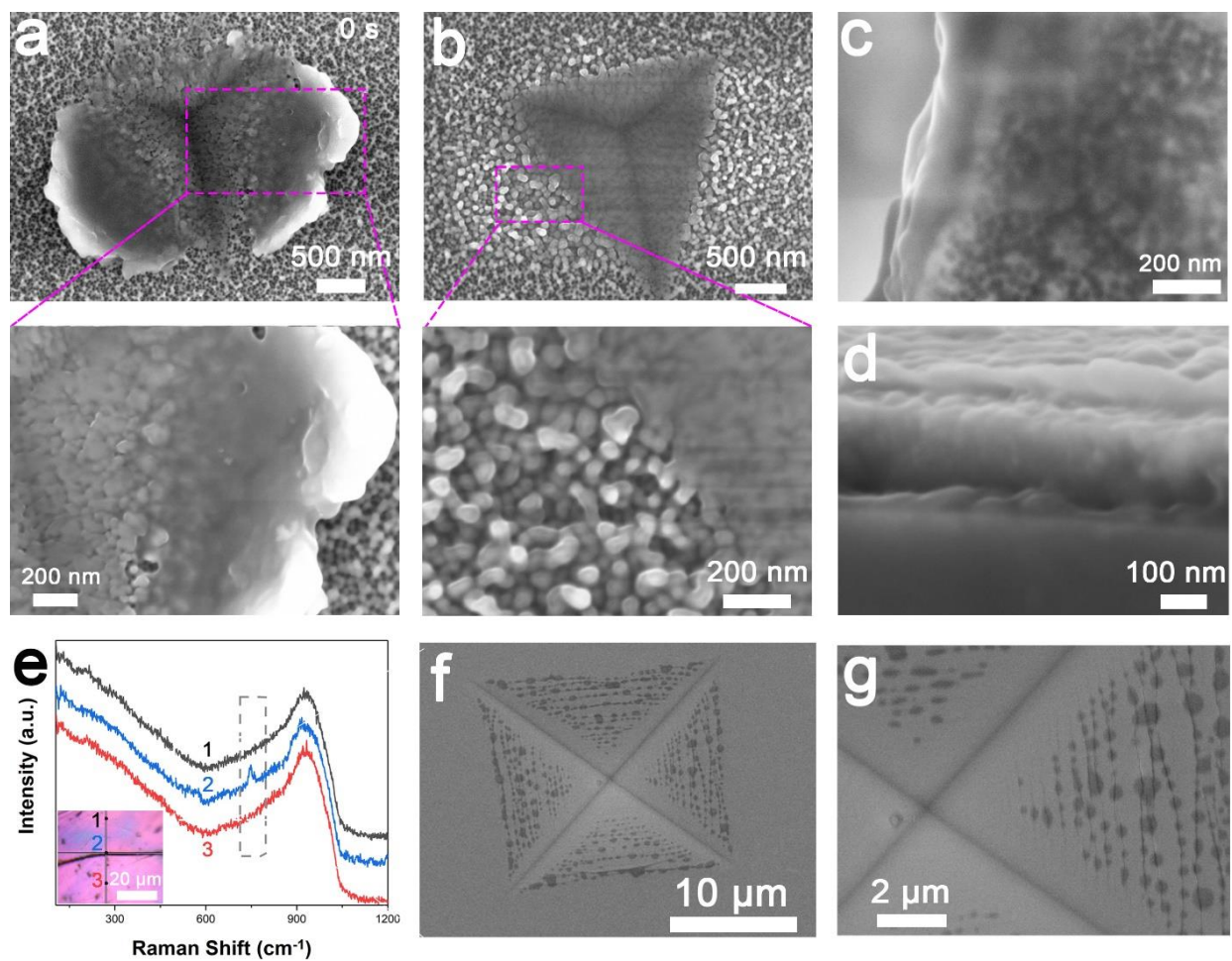

**Figure 4.** a-e) SEM images and Raman spectra taken on the $(TiAlV)O_x$ films anodized at 80 V for 2 hrs: a) after indentation pressed with a loading time of 5 s, a maximum load of 10 mN and no holding time; b) after extended e-beam irradiation at 5 keV (~ 1 min); c) top-view and d) side-view images of the fractured region (the $(TiAlV)O_x$-coated TiAlV substrate was pulled fractured into two peices); e) Raman spectra taken at the crack made on the film surface with the inset showing the scanned area under the optical microscope. f-g) SEM images of microindentation on the surface of the $(MgZnCa)O_x$ film pressed with a loading of 2 N.

## 3. Conclusion

Stress-induced liquefaction is realized on amorphous ceramic materials through a high level of atomic vacancies for the first time. With the novel liquefaction mechanisms and material design principles disclosed here, it is believed that wide-ranging ceramics of enhanced plasticity will be enabled, laying a foundation for improved applications and more scientific discoveries.

## 4. Experimental Methods

The crystalline solid solution of TiAlV was fabricated by the conventional arc melting method in the Ar atmosphere following the previously published procedure.[40] The alloy substrates were mechanically polished to a mirror finish, sonicated in acetone and ethanol, rinsed with water, and dried with nitrogen gas. Anodization was carried out in an aqueous solution of 0.1 M oxalic acid (Aldrich) using a home-made two-electrode cell with TiAlV as the anode and a Pt coil as the cathode. A constant potential (e.g., 10 - 100 V) was exerted for 2 hrs using a Keithley 2400 Sourcemeter. The anodized sample was washed with water and dried in $N_2$. The $(MgZnCa)O_x$ film was deposited on the quartz glass via RF magnetron sputtering, following the previous published procedure.[41] Briefly, the RF magnetron sputtering was carried out in the mixed gas of Ar and $O_2$ (purity > 99.99%, flow rate of 70 sccm for Ar and 10 sccm for $O_2$) with the working pressure of 0.5 Pa and sputtering power of 80 W. The substrate was kept at the room temperature and deposited with a film growth rate of approximately 9 nm/min for 20 min.

The samples were examined under a field-emission scanning electron microscope (Philips XL-30 FESEM), a transmission electron microscope (TEM, Tecnai F20) equipped with an energy-dispersive X-ray (EDX) spectroscope (Oxford INCA), an X-ray photoelectron spectroscope (XPS, VG Escalab 220i-XL, depth profiling performed with $Ar^+$ milling), and an atom probe tomography (APT) microscope (CAMECA Leap 5000 XR). Selected-area electron diffraction (SAED) patterns were recorded in the TEM. For TEM and APT measurements, the sample was sliced or carved, using a dual beam system of focused ion beam (FIB) and SEM (FEI Scios LoVac Dual Beam). X-ray diffraction (XRD) patterns were collected on a diffractometer (Rigaku SmartLab) using Cu $K_\alpha$ radiation. Mechanical properties were characterized using a nanoindenter (Hysitron TI 950 TriboIndenter), and Vickers microindenter (HXD-1000 TMC, Shanghai). Raman spectra were collected with a laser wavelength of 633 nm on a Renishaw 2000 Raman microscope calibrated using the Si peak at 520 $cm^{-1}$ as the reference.

**Supporting Information**

Supporting Information is available Online or from the author.

**Competing financial interest statement**

The authors declare no competing financial interests.

**Acknowledgements**

This work was supported by the National Key R&D Program of China (Project No. 2017YFA0204403), the Major Program of National Natural Science Foundation of China: NSFC 51590892, the Innovation and Technology Commission of HKSAR through Hong Kong Branch of National Precious Metals Material Engineering Research Centre, and the City University of Hong Kong (Project 7005077).

# Graphical abstract

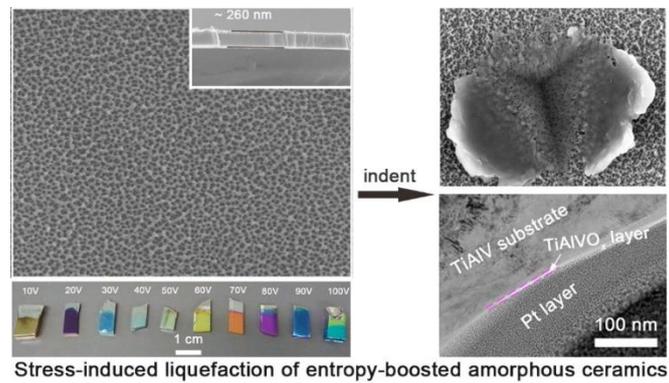

Entropy-boosted amorphous ceramics (EBACs) enable plastic deformation and high plasticity, due to a novel liquefaction mechanism.